\documentclass[twocolumn]{aastex631}

\usepackage{tablefootnote}

\shorttitle{Magnetic Field Measurements of Stellar Coronal Bases}
\shortauthors{Chen et al.}

\newcommand{\figa}[1]{Figure\,~\ref{#1}}

\newcommand{\tab}[1]{Table\,\ref{#1}}

\newcommand{\eqn}[1]{Equation\,(\ref{#1})}

\definecolor{orange}{rgb}{1,0.4,0.}
\graphicspath{{./}{figures/}}

\begin{document}

\title{Measurements of the magnetic field strengths at the bases of stellar coronae using the magnetic-field-induced transition theory}

\correspondingauthor{Hui Tian, Xianyong Bai}
\email{huitian@pku.edu.cn, xybai@bao.ac.cn}

\author{Yajie Chen}
\affiliation{School of Earth and Space Sciences, Peking University, 100871 Beijing, China}

\author{Xianyu Liu}
\affiliation{School of Earth and Space Sciences, Peking University, 100871 Beijing, China}

\author{Hui Tian}
\affiliation{School of Earth and Space Sciences, Peking University, 100871 Beijing, China}
\affiliation{Key Laboratory of Solar Activity, National Astronomical Observatories, Chinese Academy of Sciences, Beijing 100012, China}

\author{Xianyong Bai}
\affiliation{Key Laboratory of Solar Activity, National Astronomical Observatories, Chinese Academy of Sciences, Beijing 100012, China}
\affiliation{School of Astronomy and Space Science, University of Chinese Academy of Sciences, Beijing 100049, China}

\author{Meng Jin}
\affiliation{SETI Institute, 189 N Bernardo Ave suite 200, Mountain View, CA 94043, USA}

\author{Wenxian Li}
\affiliation{Key Laboratory of Solar Activity, National Astronomical Observatories, Chinese Academy of Sciences, Beijing 100012, China}

\author{Yang Yang}
\affiliation{Shanghai EBIT laboratory, Institute of Modern physics, Fudan University, Shanghai, China}
\affiliation{Key Laboratory of Nuclear Physics and Ion-beam Application (MOE), Fudan University, Shanghai 200433, China}

\author{Zihao Yang}
\affiliation{School of Earth and Space Sciences, Peking University, 100871 Beijing, China}

\author{Yuanyong Deng}
\affiliation{Key Laboratory of Solar Activity, National Astronomical Observatories, Chinese Academy of Sciences, Beijing 100012, China}
\affiliation{School of Astronomy and Space Science, University of Chinese Academy of Sciences, Beijing 100049, China}

\begin{abstract}

Measurements of the magnetic field in the stellar coronae are extremely difficult. 
Recently, it was proposed that the magnetic-field-induced transition (MIT) of the Fe~{\sc{x}} 257 {\AA} line can be used to measure the coronal magnetic field of the Sun.
We performed forward modeling with a series of global stellar magnetohydrodynamics models to investigate the possibility of extending this method to other late-type stars.
We first synthesized the emissions of several Fe~{\sc{x}} lines for each stellar model, then calculated the magnetic field strengths using the intensity ratios of Fe~{\sc{x}} 257 {\AA} to several other Fe~{\sc{x}} lines based on the MIT theory.
Finally, we compared the derived field strengths with those in the models, and concluded that this method can be used to measure at least the magnetic field strengths at the coronal bases of stars with a mean surface magnetic flux density about one order of magnitude higher than that of the Sun. Our investigation suggests the need of an extreme ultraviolet spectrometer to perform routine measurements of the stellar coronal magnetic field.  

\end{abstract}


%==================================================================
\section{Introduction} \label{sec:intro}
%==================================================================

There is increasing consensus that the magnetic activity of host star, e.g., {strong X-ray and ultraviolet emission, stellar wind}, superflares, and coronal mass ejections \citep[e.g.,][]{Baranov1990, Haisch1991, Maehara2012, Alvarado2018,Argiroffi2019, Veronig2021} could cause atmospheric loss \citep[e.g.,][]{Lammer2003, Dong2017} and surface sterilization \citep[e.g.,][]{Segura2010, Tabataba2016} of exoplanets, thus significantly affect the exoplanet habitability \citep[e.g.,][]{Linsky2019}.
The energy source of these different types of stellar activity lies mainly in the magnetic field \citep[e.g.,][]{Yokoyama1998,Pevtsov2003}. So measurements of the stellar magnetic field are important for both the stellar atmospheric characterization and {the search for extraterrestrial life}. 

Zeeman effect is often used to derive the average magnetic field strengths on stellar surfaces (photospheres) from observations of the integrated spectral line profiles \citep[e.g.,][]{Reiners2010,Reiners2012}.
Furthermore, the Zeeman-Doppler imaging technique has been applied to obtain large-scale magnetic field distributions in the photospheres of some stars \citep[e.g.,][]{Semel1989, Rosen2015}.

Measurements of the magnetic field above the stellar surfaces are very rare. 
After all, routine magnetic field measurements of the solar corona are still missing \citep[e.g.,][]{Lin2004,Liu2008,Chen2018,Chen2020,Li2017,Gary2018,Zhao2019,Zhao2021,Yang2020b,Yang2020a,Fleishman2020,Zhu2021}.
So far, only a few attempts have been made to obtain information on the stellar coronal magnetic field.
One approach is to construct the global magnetic field structures using the technique of magnetic field extrapolation based on the Zeeman-Doppler maps observed in the stellar photospheres \citep[e.g.,][]{Jardine2002,Kochukhov2002,Donati2006,Petit2008,Johnstone2014}.
However, the Zeeman-Doppler maps {can only reveal large-scale structures, and they} are subject to large uncertainties, especially when the stellar spin axes are not perpendicular to the lines of sight \citep[e.g.,][]{Lehmann2019}.
Moreover, magnetic field extrapolations usually rely on several assumptions, e.g., the magnetic field is potential, which are often not valid in stellar atmospheres \citep{Wiegelmann2021}.

Another approach is to infer the magnetic field strengths of stellar coronae from radio observations \citep[e.g.,][]{Gary1981,Mutel1985,Gudel2002}.
These observations often yielded a field strength of several hundred Gauss, which is likely for the core of active regions (ARs) or flare regions on the observed stars.
However, the radio emission mechanisms, which are critical for determination of the field strengths, are not always evident from observations.
{In addition, quiescent radio emission has not been detected on many late-type stars  \citep{White1996}.}

Recently, \citet{Li2015,Li2016} suggested that the intensity ratios of several Fe~{\sc{x}} lines can be used to measure the solar coronal magnetic field based on the magnetic-field-induced transition (MIT) theory.
The suitability of this method for measurements of the solar coronal magnetic field has been recently demonstrated by \citet{Chen2021} through a forward modeling approach.
This technique has also been successfully applied to solar extreme ultraviolet (EUV) spectral observations \citep[e.g.,][]{Si2020, Landi2020, Landi2021, Brooks2021}. The measured field strengths in solar ARs are often several hundred Gauss, which is consistent with the typical field strengths in the lower corona as derived from magnetic field extrapolations \citep{Landi2020} and magnetohydrodynamic (MHD) models \citep{Chen2021}.

Here we extend this method to the coronae of other stars through forward modeling with a series of global stellar models.
Our investigation suggests that the MIT method could be used to measure the coronal magnetic field for at least a subset of late-type stars.

%==================================================================
\section{Models and emission line synthesis} \label{sec:model}
%==================================================================

%>>>>>>>>>>>>>>>>>>>>>>>>>>>>>>>>>>>>>>>>>>>>>>>>>>>>>>>>>>>>>>>>>>>>>>>>>>>>>>
\begin{figure*} 
\centering {\includegraphics[width=\textwidth]{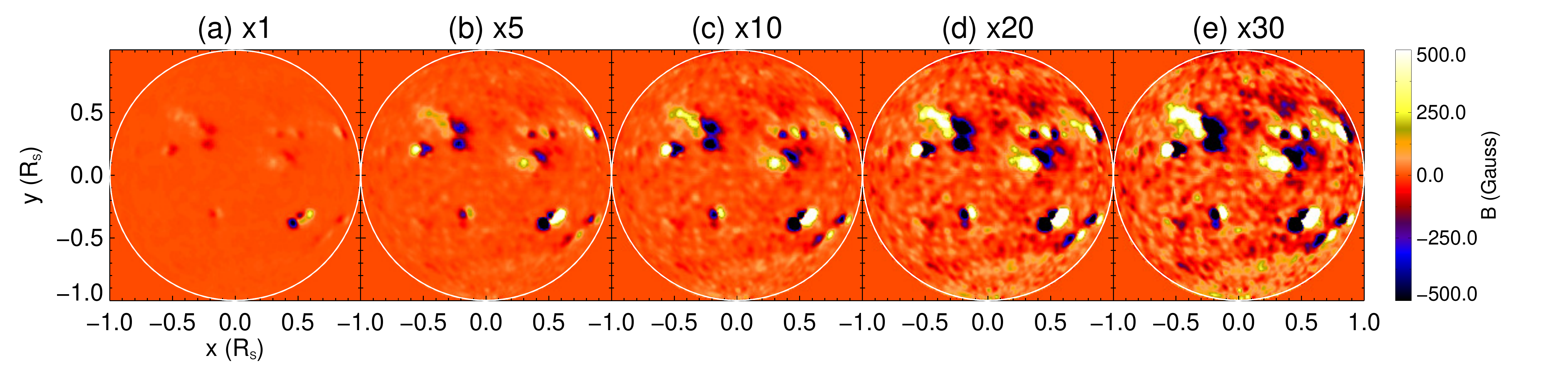}}
\caption{Radial component of the magnetic field at the stellar surface in different models. The $+x$-direction is taken as the LOS.
The white circle in each panel indicates the limb of the photosphere.
} \label{f1}
\end{figure*}
%<<<<<<<<<<<<<<<<<<<<<<<<<<<<<<<<<<<<<<<<<<<<<<<<<<<<<<<<<<<<<<<<<<<<<<<<<<<<<<

We took a series of global MHD models developed by \citet{Jin2020}.
The global coronae were constructed using the Alfv\'en Wave Solar Model \citep[AWSoM;][]{AWSoM} within the Space Weather Modeling Framework \citep[SWMF;][]{SWMF}. The AWSoM simulation domain starts from the solar upper chromosphere and extends into the corona and heliosphere. A steady-state solar wind solution is obtained with the local time stepping and second-order shock-capturing scheme. The inner boundary condition of the magnetic field is specified by the synchronous magnetogram from an evolving photospheric flux transport model \citep{Schrijver2003}. The inner boundary conditions for electron and proton temperatures $T_{e}$ and $T_{i}$ and number density $n$ are assumed to be $T_{e}=T_{i}=$ 50,000 K and $n =$ 2$\times$10$^{17}$ m$^{-3}$, respectively. The initial conditions for the solar wind plasma are specified by the Parker solution \citep{Parker1958}, while the initial magnetic field is based on the Potential Field Source Surface (PFSS) model. Alfv\'{e}n waves are driven at the inner boundary with a Poynting flux that scales with the surface magnetic field. The model includes physically consistent treatment of wave reflection, dissipation, and heat partitioning between the electrons and protons. Electron heat conduction (both collisional and collisionless) and radiative cooling are also included in the model. 

Specifically, we first constructed a steady-state global model based on the solar synchronous magnetograms {(at a resolution of 1$^{\circ}$, or $\sim$17 arcsec at the disk center)} obtained from observations of the Helioseismic and Magnetic Imager \citep[HMI,][]{Schou2012} onboard the Solar Dynamics Observatory (SDO) on 2011 Feb 15, {which is in the rising phase of solar cycle 24,} and named this solar model as the ``x1" model. More details of the numerical setups can be found in \citet{Jin2012,Jin2016} and \citet{Oran2013}.
Considering that the mean surface magnetic flux densities on many young Sun-like stars and M dwarfs are much higher than that on the Sun \citep{Reiners2012}, we then followed \citet{,Jin2020} to construct another four steady-state models by increasing the surface magnetic flux density by factors of 5, 10, 20, and 30, and referred to them as the ``x5", ``x10", ``x20", and ``x30" models, respectively.

For each model we took one snapshot of the steady-state global coronal solution for further analyses.
We defined the stellar spin axis (i.e., from the stellar center to the polar regions) as the $z$-direction and chose two orthogonal directions in the equatorial plane as the $x$- and $y$-directions.
\figa{f1} shows the radial component of the magnetic field at the stellar surfaces for a line of sight (LOS) in the $+x$-direction ($+x$ pointing towards the observer). 
The magnetic field structures on the stellar surface are the same for all models. The maximum magnetic field strength (or more precisely magnetic flux density) on the surface is $\sim$800 Gauss and the total magnetic flux is $\sim2\times10^{23}$ Mx in the x1 model.

According to the MIT theory, an external magnetic field can increase the transition rate between the Fe~{\sc{x}} 3p$^4$ 3d $^4$D$_{7/2}$ and 3p$^5$ $^2$P$_{3/2}$ levels, leading to enhanced emission of the Fe~{\sc{x}} 257.261 {\AA} spectral line \citep{Li2015,Li2016}.
Thus, the intensity ratio of the Fe~{\sc{x}} 257.261 {\AA} line to an Fe~{\sc{x}} reference line that is not sensitive to the magnetic field should change with the magnitude of the external magnetic field.
In order to perform forward modeling, we first established an atomic database of the Fe~{\sc{x}} ion.
The MIT line Fe~{\sc{x}} 257.261 {\AA} and the Fe~{\sc{x}} 257.259 {\AA} line formed through an allowed transition are too close in wavelength, and thus could not be resolved by typical EUV spectrographs with a spectral resolution ranging from several hundred to a few thousand. Because of this, we took the total intensity of these two lines as the intensity of the Fe~{\sc{x}} 257 {\AA} line.
To synthesize the coronal emissions, we followed \citet{Chen2021} and calculated the contribution functions $G(T,n_e,B)$ of the Fe~{\sc{x}} 174, 175, 177, 184 and 257 {\AA} lines as a function of electron temperature, density, and magnetic field strength \citep{Li2021}.

Stars are usually spatially unresolved when observed from the Earth. So we synthesized the spectral line emissions integrated over the whole star.
Since in most observations we can only receive radiation from the Earth-facing side of a star, the integrated emissions will vary with viewing angle.
For each model we first selected the $\pm x$-, $\pm y$-, and $\pm z$-directions as six different LOS directions.
Emissions from the far side were excluded for each LOS when calculating the intensities $I_i$ (the subscript $i$ refers to the wavelength of one Fe~{\sc{x}} line in the unit of angstrom).
For each model the intensity of each Fe~{\sc{x}} line can be calculated as
\begin{equation}
    I_i=\int_{V_1} n_e^2\cdot G_i(T,n_e,B) dV \label{eq:int}
\end{equation}
where the integration domain $V_1$ is the whole simulation box excluding the far-side region for each LOS.
%

%>>>>>>>>>>>>>>>>>>>>>>>>>>>>>>>>>>>>>>>>>>>>>>>>>>>>>>>>>>>>>>>>>>>>>>>>>>>>>>
\begin{figure*} 
\centering {\includegraphics[width=\textwidth]{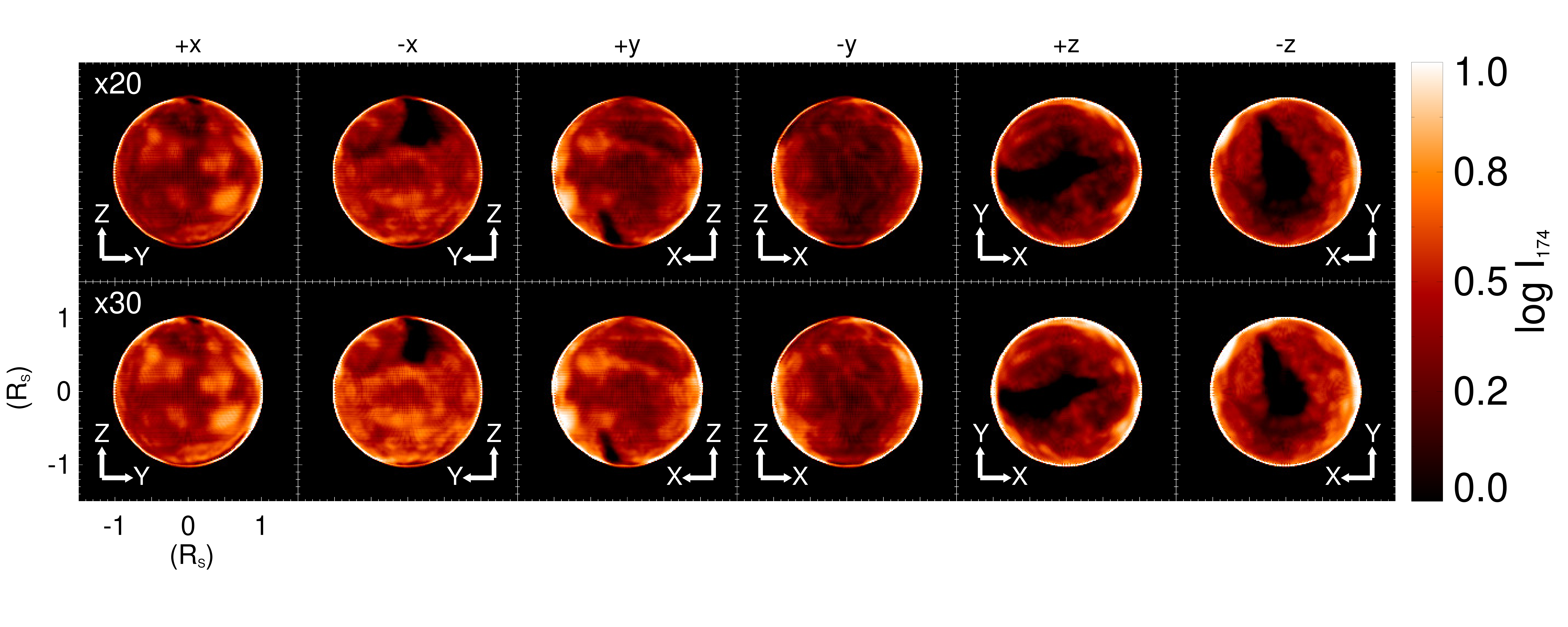}} 
\caption{Upper row: synthetic intensity images of the Fe~{\sc{x}} 174 {\AA} line for the x20 model along different LOS. 
Lower row: similar to the upper row but for the x30 model.
The $\pm z$ directions are parallel to the stellar spin axis.
The intensities are shown in logarithm scale and arbitrary unit.
} \label{f2}
\end{figure*}
%<<<<<<<<<<<<<<<<<<<<<<<<<<<<<<<<<<<<<<<<<<<<<<<<<<<<<<<<<<<<<<<<<<<<<<<<<<<<<<

%To better illustrate the results from different LOS we used in this study, 
For the purpose of understanding the different field strengths obtained from different LOS (see Section 3), we also synthesized the intensity images of the Fe~{\sc{x}} 174 {\AA} line for the x20 and x30 models along different LOS (\figa{f2}). The line intensity at every pixel of each image was calculated from \eqn{eq:int}, where the integration was performed along the corresponding LOS instead of the whole simulation box.
Since the magnetic field structures are identical in the photospheres for all the models, their corresponding coronal emission structures are also similar.
%
%Coronal holes are visible in polar regions and extend to lower latitudes.
%
Comparing the magnetograms shown in \figa{f1} (d--e) to the Fe~{\sc{x}} emission patterns presented in the first column of \figa{f2}, we can see a good correspondence, i.e., the coronal emission is enhanced in ARs where the surface magnetic field is stronger.
Due to the projection effect and limb brightening of optically thin emission, the ARs near the limb appear to be smaller and brighter in the intensity maps. The intensity maps of other Fe~{\sc{x}} lines reveal similar patterns.

%==================================================================
\section{Results and discussion} \label{sec:results}
%==================================================================

Since we used optically thin spectral lines from the Fe~{\sc{x}} ion to diagnose the magnetic field, a derived field strength should be that weighted by the intensities of these lines.
We defined an emissivity-weighted magnetic field strength from the model as:
\begin{equation}
    B_{model}=\frac{\int_{V_1} e_{174}\cdot B dV}{\int_{V_1} e_{174} dV} \label{eq2}
\end{equation}
where $e_{174}=n_e^2\cdot G_{174}$ is the emissivity of the Fe~{\sc{x}} 174 {\AA} line, and $V_1$ corresponds to the whole simulation box excluding the far-side region for a given LOS.
To evaluate the accuracy of magnetic field measurements based on the MIT theory, $B_{model}$ will be compared with the field strengths derived from intensity ratios of the synthesized Fe~{\sc{x}} lines.
%

%>>>>>>>>>>>>>>>>>>>>>>>>>>>>>>>>>>>>>>>>>>>>>>>>>>>>>>>>>>>>>>>>>>>>>>>>>>>>>>
\begin{figure} 
\centering {\includegraphics[width=80mm]{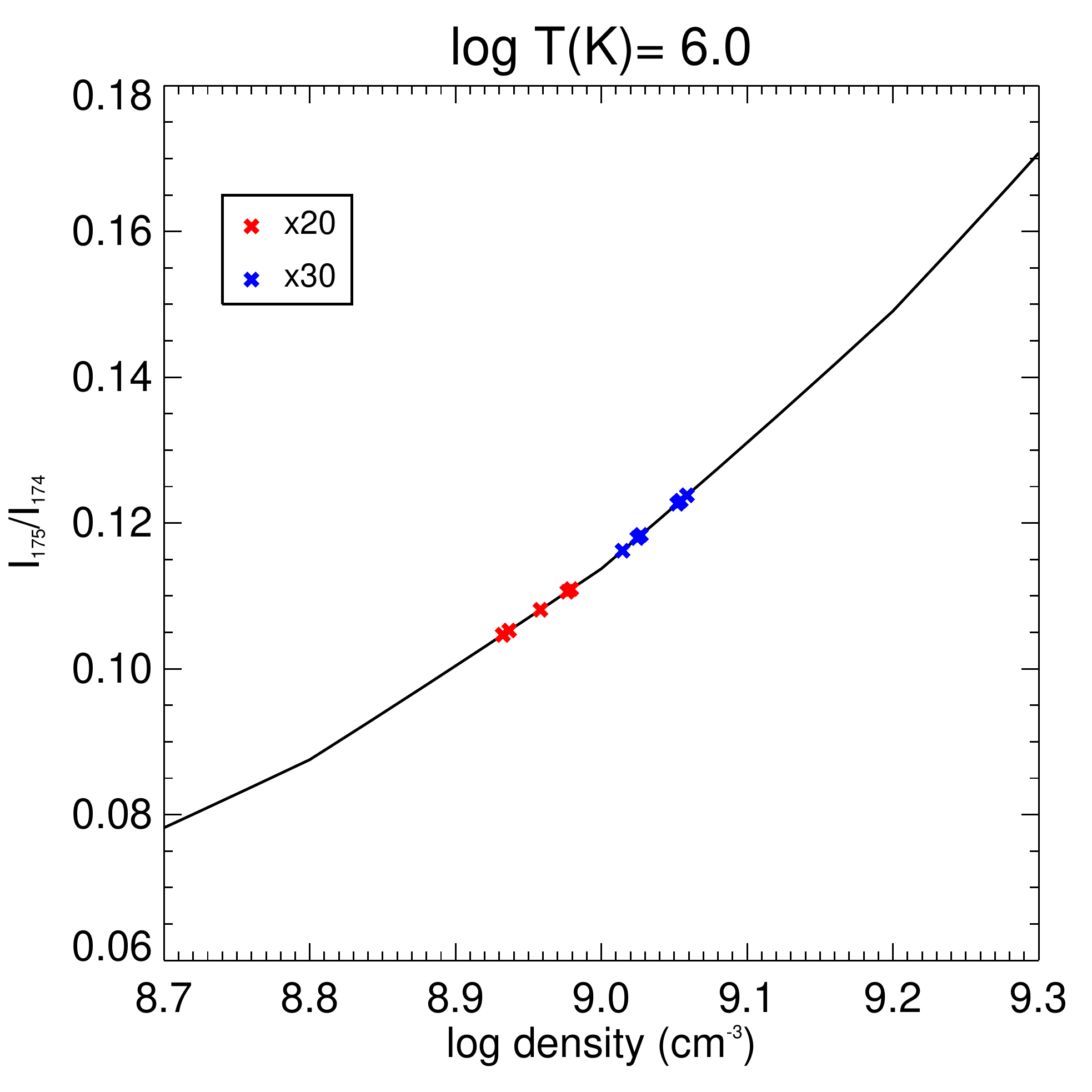}}
\caption{Intensity ratio of the Fe~{\sc{x}} 175 and 174 {\AA} line pair as a function of electron density.
The red and blue crosses indicate the derived densities (along different LOS) for the x20 and x30 models, respectively.
} \label{f3}
\end{figure}
%<<<<<<<<<<<<<<<<<<<<<<<<<<<<<<<<<<<<<<<<<<<<<<<<<<<<<<<<<<<<<<<<<<<<<<<<<<<<<<

According to the MIT theory, the magnetic field strength can be derived from the intensity ratio of Fe~{\sc{x}} 257 {\AA} to either of the Fe~{\sc{x}} 174, 177, 184 {\AA} lines \citep{Si2020,Landi2020,Chen2021}.
Since these line ratios are sensitive to both the magnetic field and electron density \citep{Li2015}, we need to calculate the density before we can obtain a magnetic field strength from the MIT method.
{For density diagnostics we took the intensity ratio of the Fe~{\sc{x}} 174/175 {\AA} line pair, which is sensitive to electron densities of  10$^{7.0}$--10$^{10.5}$ cm$^{-3}$} \citep{Brosius1998,Del_Zanna2018}.
The intensity ratio of the Fe~{\sc{x}} 257/175 {\AA} line pair is much more dependent on the density, and the error of density estimation results in enormous uncertainty of magnetic field measurements. Thus, we did not use the 257/175 {\AA} line pair to derive the magnetic field.
For a given LOS in each model, we can obtain a density value from the Fe~{\sc{x}} 174/175 {\AA} line ratio ($I_{175}/I_{174}$).
As examples, \figa{f3} presents the density diagnostic results for the x20 and x30 models.
We obtained densities of 10$^{8.9-9.0}$ and 10$^{9.0-9.1}$ cm$^{-3}$ for the x20 and x30 models, respectively.
%

%>>>>>>>>>>>>>>>>>>>>>>>>>>>>>>>>>>>>>>>>>>>>>>>>>>>>>>>>>>>>>>>>>>>>>>>>>>>>>>
\begin{table*}
\caption{Coronal magnetic field strengths estimated from the line ratios $I_{257}/I_{174}$, $I_{257}/I_{177}$, and $I_{257}/I_{184}$ for the x20 and x30 models.}
\begin{center}
\begin{tabular}{c|ccc|c||ccc|c}
\hline\hline
   & \multicolumn{4}{c||}{x20}    & \multicolumn{4}{c}{x30} \\ \cline{2-9} 
 LOS & B$_{257/174}$ & B$_{257/177}$ & B$_{257/184}$  & B$_{model}$  & B$_{257/174}$ & B$_{257/177}$ & B$_{257/184}$  & B$_{model}$  \\ \hline
$+$x & 117 ($-11\%$) & 115 ($-12\%$) & 111 ($-15\%$) & 131 & 176 ($-8\%$)  & 174 ($-9\%$) & 168 ($-13\%$) & 192 \\ \hline
$-$x & 65 (12\%)   & 64 (10\%)    & 66 (14\%)  & 58  & 101 (15\%)  & 101 (15\%) & 92 (5\%)  & 88  \\ \hline
$+$y & 116 ($-$9\%)  & 114 ($-$11\%) & 109 ($-$15\%)  & 128 & 171 ($-$8\%) & 171 ($-$8\%) & 162 ($-$13\%) & 186 \\ \hline
$-$y & 44 ($-$43\%)  & 42 ($-$45\%) & 25 ($-$68\%)   & 77  & 171 (47\%)  & 171 (47\%) & 173 (49\%) & 116 \\ \hline
$+$z & 73 ($-$21\%)  & 71 ($-$23\%) & 63 ($-$32\%)   & 92  & 189 (40\%)  & 189 (40\%) & 190 (41\%) & 135 \\ \hline
$-$z & 59 ($-$44\%)  & 58 ($-$45\%) & 48 ($-$55\%) & 106  & 122 ($-$21\%)  & 122 ($-$21\%) & 113 ($-$27\%) & 155 \\ \hline
\end{tabular}\label{tab1}
\tablecomments{The unit of the magnetic field strength is Gauss. Errors, i.e., differences between the measured field strengths and B$_{model}$, are shown in the brackets.}
\end{center}
\end{table*}
%<<<<<<<<<<<<<<<<<<<<<<<<<<<<<<<<<<<<<<<<<<<<<<<<<<<<<<<<<<<<<<<<<<<<<<<<<<<<<<

For each LOS in a model, with the estimated density and the formation temperature of the Fe~{\sc{x}} lines (10$^6$ K), we could calculate the theoretical relationship between $I_{257}/I_{i}$ (i=174, 177 or 184) and the magnetic field strength from the Fe~{\sc{x}} atomic database described in Section 2 \citep{Chen2021}. 
Using this theoretical relationship, we derived the magnetic field $B_{257/174}$, $B_{257/177}$, and $B_{257/184}$ from the synthesized Fe~{\sc{x}} line ratios $I_{257}/I_{174}$, $I_{257}/I_{177}$, and $I_{257}/I_{184}$, respectively.
The results for the x20 and x30 models are summarized in \tab{tab1}.
The magnetic field strengths in the models ($B_{model}$) and errors of the magnetic field measurements $(B-B_{model})/B_{model}$ are also included in \tab{tab1}.

%>>>>>>>>>>>>>>>>>>>>>>>>>>>>>>>>>>>>>>>>>>>>>>>>>>>>>>>>>>>>>>>>>>>>>>>>>>>>>>
\begin{figure*} 
\centering {\includegraphics[width=15cm]{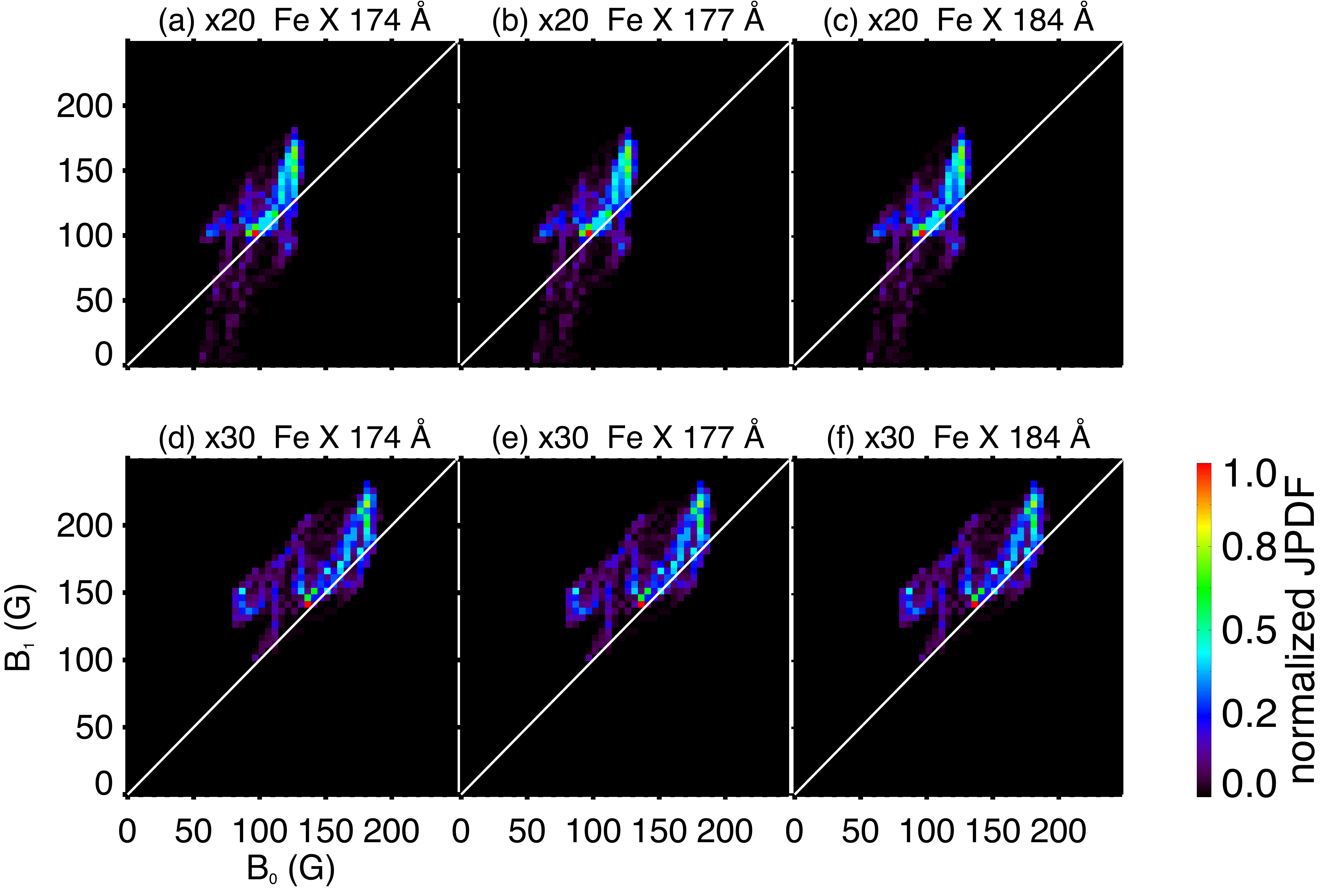}} 
\caption{{The joint probability density function (JPDF) of coronal magnetic field strengths in the models ($B_0$) and inferred ($B_1$) from the intensity ratios of the Fe~{\sc{x}} 257/174 {\AA} (left panels), 257/177 {\AA}  (middle panels), and 257/184 {\AA} (right panels) line pairs based on the MIT method taking different LOS directions. Panels (a)-(c) correspond to the x20 model, and panels (d)-(f) correspond to the x30 model. }
} \label{f2b}
\end{figure*}
%<<<<<<<<<<<<<<<<<<<<<<<<<<<<<<<<<<<<<<<<<<<<<<<<<<<<<<<<<<<<<<<<<<<<<<<<<<<<<<

In the x20 and x30 models, $B_{model}$ changes from several tens to more than one hundred Gauss when measured along different LOS.
The number of observed ARs changes with LOS as shown in \figa{f2}, and the magnetic field strengths in ARs are usually much larger than those in the quiet regions.
Thus, the magnetic field strengths derived from different LOS in the same model can be different.
Then we compared the $B_{model}$ values to the magnetic field strengths derived using the Fe~{\sc{x}} 257/174, 257/177, and 257/184 {\AA} line pairs.
We found that all the three Fe~{\sc{x}} line pairs can provide reasonably accurate estimates of the coronal magnetic field in the models. 
The differences between the measured field strengths and B$_{model}$ values are mostly smaller than $\sim$50$\%$, demonstrating that the MIT method can be extended to some late-type stars with strong surface magnetic field fluxes.
{Furthermore, we selected more LOS directions, i.e., every 5$^{\circ}$ for both the inclination and azimuth, and derived the magnetic field strengths in the model following \eqn{eq2} ($B_0$) and inferred from the MIT method ($B_1$) for each LOS directions. The comparisons between $B_0$ and $B_1$ are shown in \figa{f2b}, which do not alter the conclusion mentioned above. 
So the selection of different LOS directions (thus inclination of the stellar rotation axis with respect to the LOS) does not affect the validation of the MIT method.
In addition, we noticed some vertical features in \figa{f2b}. These features do not change much if we use LOS directions with one inclination, suggesting that they are not related to the inclination.
}

The average coronal temperature in the solar model is $\sim$10$^{6.0}$ K.
The average coronal temperatures in the x20 and x30 models reach up to $\sim$10$^{6.5}$ K \citep{Jin2020}.
Considering that the formation temperatures of the Fe~{\sc{x}} lines peak at $\sim$10$^{6.0}$ K, these Fe~{\sc{x}} lines are likely formed in the upper transition region or coronal base of these models.
So the derived field strengths correspond to the magnetic field at the very bottom parts of the stellar coronae.
This also explains why the structures of the Fe~{\sc{x}} line shown in \figa{f2} do not extend much above the limb.

%>>>>>>>>>>>>>>>>>>>>>>>>>>>>>>>>>>>>>>>>>>>>>>>>>>>>>>>>>>>>>>>>>>>>>>>>>>>>>>
\begin{figure}
\centering {\includegraphics[width=0.47\textwidth]{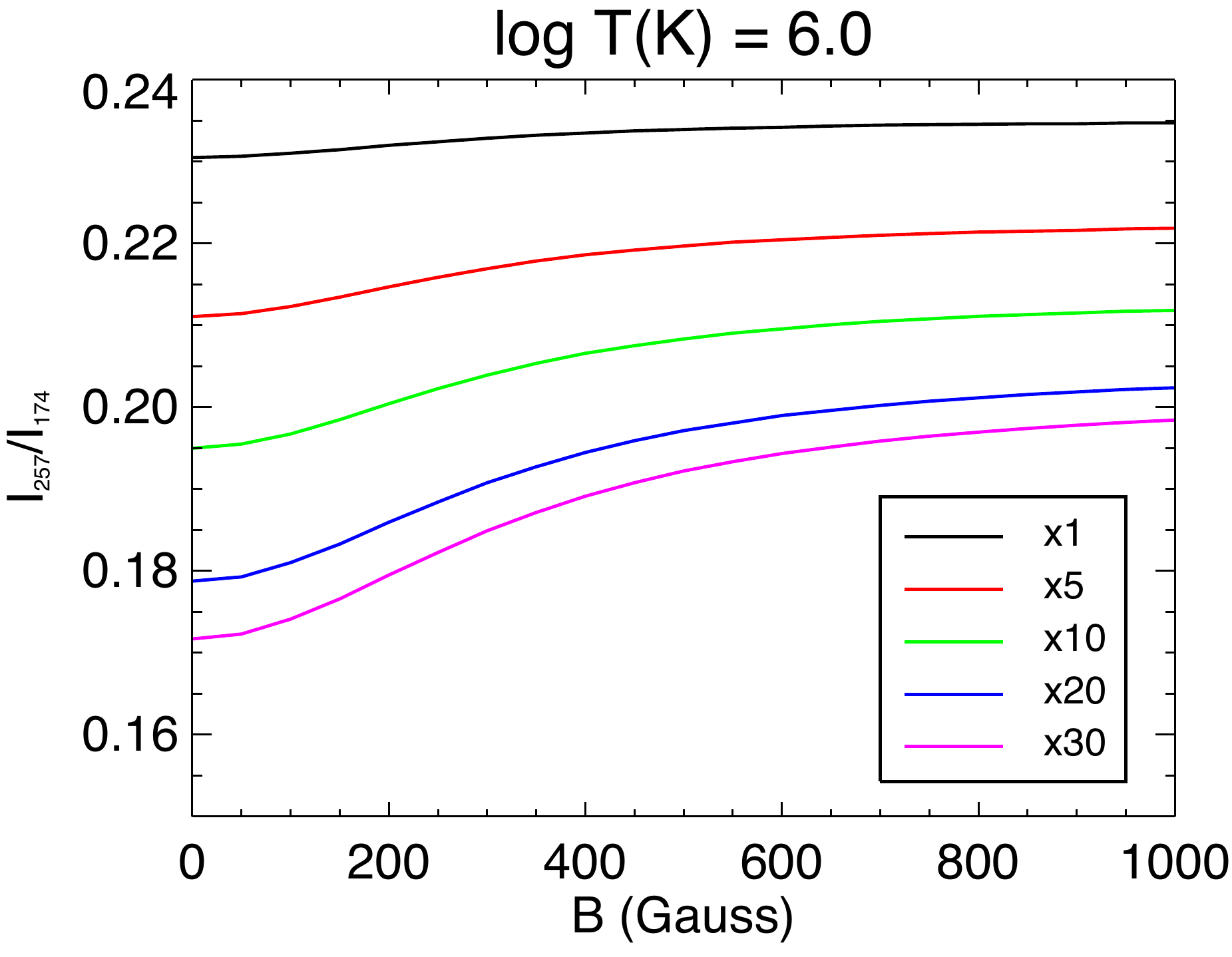}} 
\caption{The Fe~{\sc{x}} 257/174 {\AA} line ratio as a function of the magnetic field strength at the average coronal density ($n_0$) in each model.  The black, red, green, blue, and purple curves represent results for the x1, x5, x10, x20, x30 models, respectively.
} \label{f6}
\end{figure}
%<<<<<<<<<<<<<<<<<<<<<<<<<<<<<<<<<<<<<<<<<<<<<<<<<<<<<<<<<<<<<<<<<<<<<<<<<<<<<<

However, we found that most values of $I_{257}/I_{i}$ in the x1, x5, and x10 models do not fall into the ranges predicted by the MIT theory. We have investigated the reasons why this technique fails for the x1, x5, and x10 models. We defined the coronal density in each model as,
\begin{equation}
    n_0=\frac{\int_V e_{174}\cdot n_e dV}{\int_V e_{174} dV} \label{eq:em_n}
\end{equation}
where the integration volumn is the whole calculation domain, without consideration of different LOS.
We obtained $n_0$ values of 10$^{8.13}$, 10$^{8.58}$, 10$^{8.80}$, 10$^{8.99}$, and 10$^{9.07}$ cm$^{-3}$ for the x1, x5, x10, x20, and x30 models, respectively.
\figa{f6} shows the Fe~{\sc{x}} 257 and 174 {\AA} line ratio as a function of the magnetic field strength, taking the density of $n_0$ in each model.
It is obvious that the magnetic field sensitivity of $I_{257}/I_{174}$ becomes worse as the density decreases.
For the solar model, $I_{257}/I_{174}$ hardly changes with the magnetic field.
Similar results have also been revealed from the Fe~{\sc{x}} 257/177 and 257/184 {\AA} line ratios. 
\figa{f6} also shows that a density variation can lead to a change in the range of ratios that correspond to the field strengths of 0--1000 Gauss. Because the ranges of ratios are narrower in the low-density x1, x5, and x10 models, a small deviation of the estimated density from the density in the model can cause a given value of synthesized line ratio to fall outside the range of ratios predicted using the estimated density. As a result, a magnetic field strength cannot be inferred for these low-density models. 

We noticed that the average coronal density at the Fe~{\sc{x}} line formation heights of the solar model ($\sim$10$^{8.1}$ cm$^{-3}$) is about one order of magnitude lower than the typical AR density in the lower corona, i.e., $\sim$10$^{9.2}$ cm$^{-3}$ as estimated from intensity ratios of density-sensitive Fe~{\sc{xii}} line pairs in on-disk observations \citep{Tian2012}. It also appears to be a few times lower than the typical quiet-Sun density estimated from on-disk observations of density-sensitive Fe~{\sc{xii}} and Fe~{\sc{xiii}} line pairs \citep{Dere2007}. 
Several reasons could contribute to this issue: 
(1) Because of the fixed density at the inner boundary of the simulation box, the adopted model lacks the process of chromospheric evaporation resulting from conduction of coronal heating, leading to less plasma in the corona of ARs \citep{Oran2013}; 
(2) The spatial resolution of the global model is low, so that small-scale dense structures in the lower corona of both the quiet Sun and ARs cannot be well resolved; 
(3) The temperature in the solar model reaches $\sim$1 MK at larger heights as compared to observations \citep{Oran2013}, meaning that in the model the Fe~{\sc{x}} lines are formed at larger heights where the density is lower. 
Consequently, the average density at the Fe~{\sc{x}} line formation heights of the solar model is lower than that of the real solar corona. Similarly, the average densities in the lower coronae of all the stellar models are likely lower than those of the real stellar coronae. Since the magnetic field sensitivity of the $I_{257}/I_{i}$ ratio increases with density, we expect that our technique can be extended further to stars with a surface magnetic flux smaller than that in the x20 model. In the near future we plan to quantitatively verify this using stellar models with more realistic coronal densities. 

{We realized that the scaled magnetograms may not capture the full complexity of the magnetic field on some stars, especially the strong quadrupolar/octopolar components and surface toroidal fields that are present on some young/rapidly rotating stars. However, we think that for the first step of a proof-of-concept investigation, our usage of the scaled magnetograms is still reasonable. This is because the complexity (thus detailed structures) of the stellar magnetic field should not affect the suitability of our technique, as the line ratios of the MIT line and other reference lines are only sensitive to the coronal temperature, density, and magnetic field strength. The average magnetic field strengths (magnetic flux densities) measured on the photospheres of some stars could be up to three orders of magnitude larger than that of the Sun \citep[e.g.,][]{Pevtsov2003,Kochukhov2021}. The larger flux density might be explained by either of the following two scenarios or a combination of the two: 1) a stronger field in regions with typical solar AR sizes, 2) a larger surface area with a field strength of typical solar AR field strength. In this study, we chose the first scenario since we can make use of the directly observed solar magnetograms. In our x30 model, the maximum field strength is $\sim$25 kG, which should not be too unrealistic if we consider the fact that even the solar photospheric magnetic field can reach $\sim$7.5 kG \citep{vanNoort2013}. The second scenario would involve larger sizes of individual ARs or a larger number of ARs. In this situation a photospheric magnetogram at the model inner boundary cannot be easily constructed from actual observations. Nevertheless, in the near future we plan to construct stellar magnetograms through other approaches, e.g., using stellar flux transport models \citep{Schrijver2020}, and explore the suitability of coronal magnetic field diagnostics with the MIT method.}

It is worth mentioning that the interstellar medium could cause absorption of the EUV spectral lines of the Fe~{\sc{x}} ions we used in this study \citep{Rumph1994}.
However, for many stars in the solar neighborhood, the interstellar absorption of these Fe~{\sc{x}} lines is weak or could be evaluated based on models \citep{France2019}.
So as long as these Fe~{\sc{x}} lines are measured through a future EUV spectrograph, the MIT method can be used to infer the magnetic field strengths at the coronal bases of many nearby stars.

%==================================================================
\section{Summary} \label{sec:summary}
%==================================================================

We have proposed a new method to carry out stellar coronal magnetic field measurements using the magnetic-field-induced transition of the Fe~{\sc{x}} 257 {\AA} line.
To investigate the feasibility of this method, we constructed a series of steady-state global stellar models by multiplying the surface magnetic flux density of the real solar observations by factors of 1, 5, 10, 20, and 30 to obtain the x1, x5, x10, x20, and x30 models, respectively.
We synthesized the emissions of the Fe~{\sc{x}} 174, 175, 177, 184, and 257 {\AA} lines, and then derived the coronal electron densities and magnetic field strengths successively using the intensity ratios of different Fe~{\sc{x}} line pairs for different LOS directions.
After that, we compared the derived coronal magnetic field strengths to those in the models.

We found that all the Fe~{\sc{x}} 257/174, 257/177, and 257/184 {\AA} line ratios can provide reasonably accurate magnetic field measurements at the  coronal bases for the x20 and x30 models.
Further analyses suggest that the low densities at the lower coronae may have resulted in the failure of the application of this technique for the x1, x5, and x10 models.
Our investigation demonstrates that the MIT method can be used to measure the magnetic field strengths at the coronal bases of at least some nearby M dwarfs and young Sun-like stars whose magnetic field strengths are more than one order of magnitude stronger than that of the Sun. To achieve routine measurements of the coronal magnetic field of these stars, an EUV spectrometer that can simultaneously measure at least the Fe~{\sc{x}} 174, 175, and 257 {\AA} lines is highly desired.

\begin{acknowledgments}
This  work  is  supported  by  NSFC  grants 11825301, 11790304, 11704076, 12073004 and U1732140, the Strategic Priority Research Program of CAS (grant no.  XDA17040507) and grant 1916321TS00103201.
\end{acknowledgments}

\bibliography{mitstellar}{}
\bibliographystyle{aasjournal}

\end{document}